\documentclass[a4paper]{iag}
\usepackage[dvips]{graphics}
\usepackage{epsfig,graphicx}

\usepackage{natbib_2}

\begin{document}

\title{Potential Capabilities of Lunar Laser Ranging for Geodesy and Relativity}

\author{ J\"urgen M\"uller\\
Institut f\"{u}r Erdmessung, University of Hannover, Schneiderberg 50, 30167 Hannover, Germany\\
mueller@ife.uni-hannover.de\\
James G.\ Williams, Slava G.\ Turyshev\\
Jet Propulsion Laboratory, 4800 Oak Grove Drive, Pasadena, CA 91109, USA\\
Peter J. Shelus\\
University of Texas at Austin, Center for Space Research, 3925 W. Braker Lane, Austin, TX 78759, USA}

\maketitle\thispagestyle{empty}

\abstract{
Lunar Laser Ranging (LLR), which has been carried out for more than 35 years, is used to determine many parameters within the Earth-Moon system. This includes coordinates of terrestrial ranging stations and that of lunar retro-reflectors, as well as lunar orbit, gravity field, and its tidal acceleration.  LLR data analysis also performs a number of gravitational physics experiments such as test of the equivalence principle, search for time variation of the gravitational constant, and determines value of several metric gravity parameters.  These gravitational physics parameters cause both secular and periodic effects on the lunar orbit that are detectable with LLR.  Furthermore, LLR contributes to the determination of 
Earth orientation parameters (EOP) such as nutation, precession (including relativistic precession), polar motion, 
and UT1. The corresponding LLR EOP series is three decades long. LLR can be used for the realization of both the terrestrial and selenocentric reference frames.  The realization of a dynamically defined inertial reference frame, in contrast to the kinematically realized frame of VLBI, offers new possibilities for mutual cross-checking and 
confirmation. Finally, LLR also investigates the processes related to the Moon's interior dynamics. 

Here, we review the LLR technique focusing on its impact on Geodesy and Relativity. We discuss the modern observational accuracy and the level of existing LLR modeling. We present the near-term objectives and emphasize improvements needed to fully utilize the scientific potential of LLR. 
}


\noindent\rule[0mm]{\linewidth}{0.2mm}

\section{Motivation}
Being one of the first space geodetic techniques, lunar laser ranging (LLR) has routinely provided observations for more than 35 years.  The LLR data are collected as normal points, i.e. the combination of round trip light times of lunar returns obtained over a short time span of 10 to 20 minutes. Out of $\approx 10^{19}$ photons sent per pulse by the transmitter, less than 1 is statistically detected at the receiver \cite{wil96}; this is because of the combination of several factors, namely energy loss (i.e. $1/R^4$ law), atmospherical extinction and geometric reasons (rather small telescope apertures and reflector areas). Moreover, the detection of real lunar returns is rather difficult as dedicated data filtering (spatially, temporally and spectrally) is required. These conditions are the main reason, why only a few observatories worldwide are capable of laser ranging to the Moon.

Observations began shortly after the first Apollo 11 manned mission to the Moon in 1969 which deployed a passive retro-reflector on its surface. Two American and two French-built reflector arrays (transported by Soviet spacecraft) followed until 1973. Since then over 16,000 LLR 
measurements have by now been made of the distance between Earth observatories and lunar reflectors. Most LLR data have been collected by a site operated by the Observatoire de la C\^ote d'Azur (OCA), France.  The transmitter/receiver used by OCA is a 1.5m alt-az Ritchey-Chr\'etien reflecting telescope. The mount and control electronics insure blind tracking on a lunar feature at the 1 arcsec level for 10 minutes. The OCA station uses a neodymium-YAG laser, emitting a train of pulses, each with a width of several tens of picoseconds).  LLR station at the McDonald Observatory in Texas, USA is another major provider of the LLR data.  The McDonald Laser Ranging Station (MLRS) is built around a computer-controlled 0.76-m x-y mounted Cassegrain/Coud\'e reflecting telescope and a short pulse, frequency doubled, 532-nm, neodymium-YAG laser with appropriate computer, electronic, meteorological, and timing interfaces.  Until 1990, Haleakala laser ranging station on the island of Maui (Hawaii, USA) contributed to LLR activities with its 0.40~m telescope. Single lunar returns are available from Orroral laser ranging station in  Australia (closed 1 November 1998) and the Wettzell Laser Ranging System (0.75~m) in Germany.  Other modern stations have demonstrated lunar capability, e.g. the Matera Laser Ranging Station (0.50~m) in Italy and Hartebeesthook Observatory (0.762~m) in South Africa. A new site with lunar capability is currently being built at the Apache Point Observatory (New Mexico, USA) around a 3.5~m telescope.  This station, called APOLLO, is designed for mm accuracy ranging (Williams et al. 2004b). 

Today MLRS and OCA are the only currently operational LLR sites achieving a typical range precision of 18-25 mm. Fig.~1 shows the number of LLR normal points per year since 1970. As shown in Fig.~2, the range data have not been accumulated uniformly; substantial variations in data density 
exist as a function of synodic angle~$D$, these phase angles are represented by 36 bins of 10 degree width. 
In Fig.~2, data gaps are seen near new Moon (0 and 360 degrees) and full Moon (180 degrees) phases.
 The properties of this data distribution are a consequence of operational restrictions, such as difficulties to operate near the bright sun in daylight (i.e. new Moon) or of high background solar illumination noise (i.e. full Moon).   

\begin{figure}[thb]
  \begin{center}
  \epsfig{width=7.4cm,     file=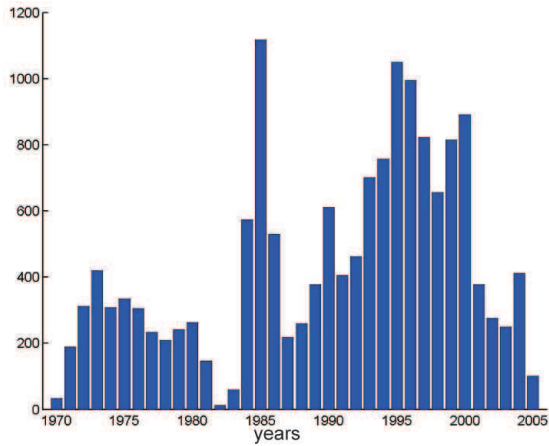}
  \end{center}\vskip -5pt
  \caption{Lunar observations per year, 1970 - 2005.} \label{annobs}
 \end{figure}
 
\begin{figure}[thb]
  \begin{center}
  \epsfig{width=7.4cm,     file=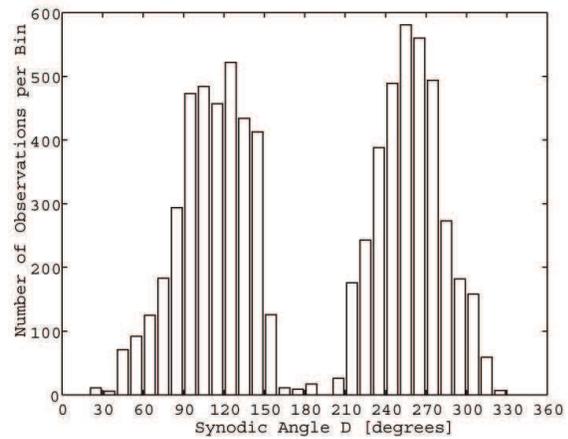}
  \end{center}\vskip -5pt
  \caption{Data distribution as a function of the synodic angle $D$.} \label{synmo}
 \end{figure}

While measurement precision for all model parameters benefit from the ever-increasing improvement in precision of individual range measurements (which now is at the few cm level, see also Fig.\ 3), some parameters of scientific interest, such as time variation of Newton's coupling parameter $\dot G/G$ or precession rate of lunar 
perigee, particularly benefit from the long time period (35 years and growing) of range measurements. 


\begin{figure}[thb]
  \begin{center}
  \epsfig{width=7.4cm,     file=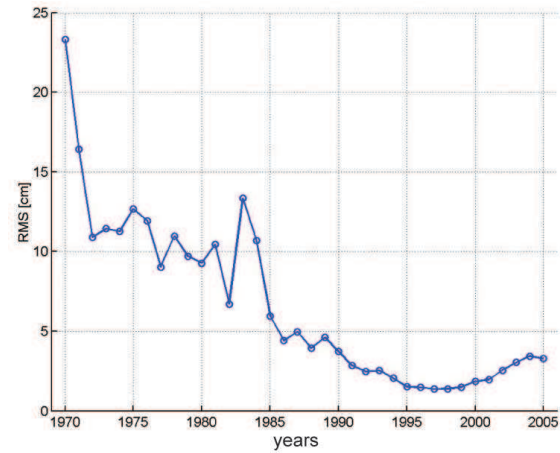}
  \end{center}
  \caption{Weighted residuals (observed-computed Earth-Moon distance) annually averaged.} \label{annrms}
 \end{figure}
 
In the 1970s LLR was an early space technique for determining 
Earth orientation parameters (EOP). EOP data from LLR are computed for those nights where sufficient data were available (approximately 2000 nights over 35 years). Although, the other space geodetic techniques (i.e., SLR, VLBI) dominate since the 80s, today LLR still delivers very competitive results, and because of large improvements in ranging precision (30 cm in 1969 to 2~cm today), it now serves as one of the strongest tools in the solar system for testing general relativity. Moreover, parameters such as the station coordinates and velocities 
contributed to the International Terrestrial Reference Frame ITRF2000, EOP quantities were used in combined 
solutions of the International Earth Rotation and Reference Systems Service IERS ($\sigma$~=~0.5~mas).

\section{LLR Model}

The existing LLR model has been developed to compute the LLR observables --- the round trip travel times of laser pulses between stations on the Earth and passive reflectors on the Moon (see e.g.\ M\"uller et al.\ 1996, M\"uller  and
Nordtvedt 1998 or M\"uller 2000, 2001, M\"uller and Tesmer 2002, Williams et al.\ 2005b and the references therein). The model is fully relativistic and is complete up to first post-Newtonian ($1/c^2$) level; it uses the  Einstein's general theory of relativity -- the standard theory of gravity. The modeling of the relativistic parts is much more 
challenging than, e.g., in SLR, because the relativistic corrections increase the farther the distance becomes.
The modeling of the `classical' parts has been set up according to IERS Conventions (IERS 2003), but it is restricted to the 1 cm level.  Based upon this model, two groups of parameters (170 in total) are determined by a weighted least-squares fit of the observations. 
The first group comprised from the so-called  1Newtonian' parameters such as
\begin{itemize}
\item[--] geocentric coordinates of three Earth-based LLR stations and their velocities;
           
\item[--] a set of EOPs (luni-solar precession constant, nutation coefficients of the 18.6 years period, Earth's rotation UT0 and variation of latitude by polar motion);

\item[--] selenocentric coordinates of four retro-reflectors;

\item[--] rotation of the Moon at one initial epoch (physical librations);

\item[--] orbit (position and velocity) of the Moon at this epoch;

\item[--] orbit of the Earth-Moon system about the Sun at one epoch;

\item[--] mass of the Earth-Moon system times the gravitational constant;

\item[--] the lowest mass multipole moments of the Moon;

\item[--] lunar Love number and a rotational energy dissipation parameter;

\item[--] lag angle indicating the lunar tidal acceleration responsible for the increase of the Earth-Moon distance
(about 3.8 cm/yr), the increase in the lunar orbit period and the slowdown of Earth's angular velocity.
\end{itemize}
    
The second group of parameters used to perform LLR tests of plausible modifications of general theory of relativity (these parameter values for general relativity are given in parentheses):
\begin{itemize} 
\item[--] geodetic de {Sitter} precession ${\Omega}_{\rm dS}$  of 
the lunar orbit ($\simeq 1.92 "/cy$);

\item[--] space-curvature parameter $\gamma$ (= 1) and non-linearity parameter $\beta$ (= 1);

\item[--] time variation of the gravitational coupling parameter $\dot G/G$ (= 0  $\rm yr^{-1}$) which is important
for the unification of the fundamental interactions;

\item[--] strong equivalence principle (EP) parameter, which for metric theories is $\eta=4\beta-3-\gamma$ (= 0);

\item[--] EP-violating coupling of normal matter to 'dark matter' at the galactic center;

\item[--] coupling constant $\alpha$~(=~0) of Yukawa potential for the Earth-Moon distance which corresponds to a test of
Newton's inverse square law;

\item[--] combination of parameters $\zeta_1 - \zeta_0 - 1$ (= 0) derived in the Mansouri and Sexl (1977) formalism indicating a violation of special relativity (there: 
Lorentz contraction parameter $\zeta_1 = 1/2$, time dilation parameter $\zeta_0 = - 1/2$);

\item[--] $\alpha_1$ (= 0) and $\alpha_2$ (= 0) which parameterize 'preferred frame' effects in metric gravity.
\end{itemize}

Most relativistic effects produce periodic perturbations of the Earth-Moon range
$$
\Delta r_{EM} = \sum_{i=1}^n A_i \cos (\omega_i \Delta t + \Phi_i). \eqno(1)
$$
$A_i,\; \omega_i,$ and $\phi_i$ are the amplitudes, frequencies, and phases, respectively, of 
the various perturbations. Some example periods of perturbations important for the measurement of various 
parameters are given in Table~1.
{Note: the 
designations should not be used as formulae for the computation of the corresponding periods, e.g.\ the period 
`sidereal-2$\cdot$annual' has to be calculated as $1/(1/27.32^d - 2/365.25^d) \approx 32.13^d$.\\
`secular + emerging periodic' means
the changing orbital frequencies induced by $\dot G/G$ are starting to become better signals than the 
secular rate of change of the Earth-Moon range in LLR.}

\begin{figure}[tbh]
  \begin{center}
  \epsfig{width=7.4cm,     file=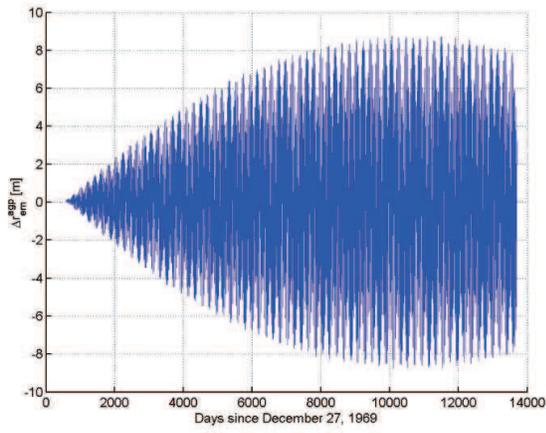}
  \end{center}\vskip-5pt
  \caption{Sensitivity of LLR with respect to $\dot G / G$ assuming $\Delta \dot G / G = 8\cdot 10^{-13}\ {\rm yr}^{-1}$.} \label{gpar}
 \end{figure}

\begin{figure}[thb]
  \begin{center}
  \epsfig{width=7.4cm,     file=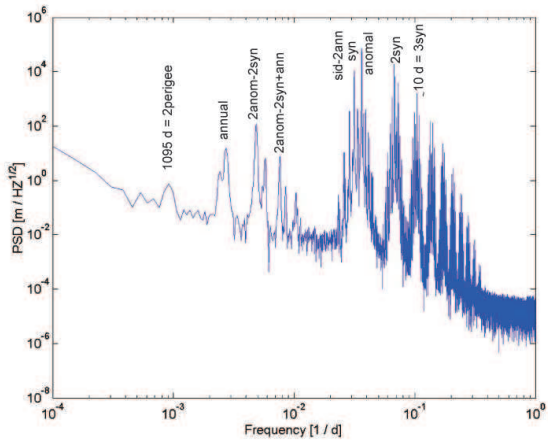}
  \end{center}\vskip-5pt
  \caption{Power spectrum of the effect of $\dot G / G$ in the Earth-Moon distance assuming $\Delta \dot G / G 
  = 8\cdot 10^{-13}\ {\rm yr}^{-1}$.}  \label{gspec}
 \end{figure}

\begin{figure}[thb]
  \begin{center}
  \epsfig{width=7.4cm,     file=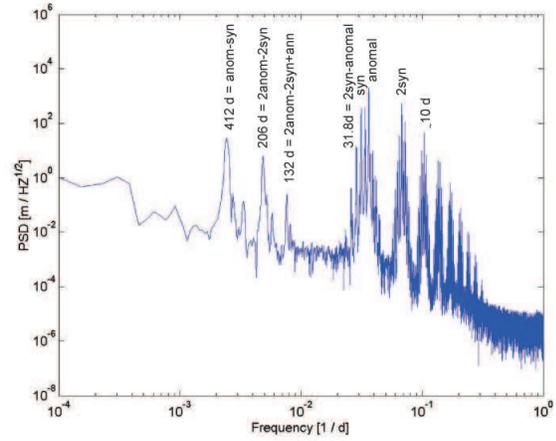}
  \end{center}\vskip-5pt
  \caption{Power spectrum of a possible equivalence principle violation  assuming $\Delta (m_G / m_I) \approx 10^{-13}$.} \label{etaspec}
 \end{figure}

Fig.~4 represents the sensitivity of the Earth-Moon distance with respect to a possible temporal variation of the gravitational  constant in the order of $8\cdot 10^{-13}$, the present accuracy of that parameter.
It seems as if perturbations of up to 9 meters are still caused, but this range (compared to the ranging accuracy at the cm level) can not fully be exploited, because the lunar tidal acceleration perturbation is similar. The largest periods for $\dot G / G$ are shown in Fig.~5 and for the EP-parameter in Fig.~6. Obviously many periods are affected simultaneously, because the perturbations, even if caused by a single beat period only (e.g. the synodic month for $\eta$), change the whole lunar orbit (and rotation) and therefore excite further frequencies.
Nevertheless these properties can be used to identify and separate the different effects and to determine corresponding parameters (note that relativistic phenomena show up with typical periods). Other effects, like the asteroids cause similar orbit perturbations (i.e. with the same frequencies) but with different amplitudes, so that they can be distinguished and separated from the effects investigated.

 \begin{table}[bht]
\caption{Typical periods of some relativistic quantities, taken from M\"uller et al.\ (1999).}
\begin{center}
\begin{tabular}{|c|c|}
\hline
Parameter & Typical Periods \\ \hline\hline
$\eta$ &  synodic (29$^d$12$^h$44$^m$2.9$^s$)\\ \hline
$\zeta_1$-$\zeta_0$-1 & annual (365.25$^d$)\\ \hline
$\delta g_{\rm galactic}$ & sidereal (27$^d$7$^h$43$^m$11.5$^s$)\\ \hline
$\alpha_1$ & sidereal, annual, sidereal-2$\cdot$annual,\\
           & anomal.\ (27$^d$13$^h$18$^m$33.2$^s$)$\pm$annual, synodic\\ \hline
$\alpha_2$ & 2$\cdot$sidereal, 2$\cdot$sidereal-anomal., nodal (6798$^d$) \\ \hline
$\dot G/G$ & secular + emerging periodic  \\ \hline
\end{tabular}
\end{center}
\end{table}

\section{Results}
The global adjustment of the model by least-squares-fit procedures gives improved values for the estimated parameters and their formal standard errors, while consideration of parameter correlations obtained from the covariance analysis and of model limitations lead to more 'realistic' errors. Incompletely modeled solid Earth tides, ocean loading or 
geocenter motion, and uncertainties in values of fixed model parameters have to be considered in those estimations.
 For the temporal variation of the gravitational constant, $\dot G / G =  (6\pm 8)\cdot 10^{-13}$ has been 
 obtained, where the formal standard deviation has been scaled by a factor 3  to yield the given value. This parameter benefits most from the long time span of LLR data and has experienced  the biggest improvement over the past years (cf.\ M\"uller et al.\ 1999). In contrast, the EP-parameter $\eta$ $\left( =(6 \pm 7) \cdot 10^{-4}\right)$ benefits most from highest accuracy over a sufficient long time span (e.g.\ one year) and a good data coverage over the synodic month, as far as possible. Its improvement was not so big, as the 
LLR RMS residuals increased a little bit in the past years, compare Fig.\ 3. The reason for that increase is not completely understood and has to be investigated further.
In combination with the recent value of the space-curvature parameter $\gamma_{\rm Cassini}$ $\left(\gamma -1 = (2.1 \pm 2.3) \cdot 10^{-5}\right)$ derived from Doppler measurements to the Cassini spacecraft  (Bertotti et al.\ 2003),
the non-linearity parameter $\beta$ can be determined by applying the relationship 
$\eta=4\beta-3-\gamma_{\rm Cassini}$. One obtains
$\beta - 1 = (1.5 \pm 1.8) \cdot 10^{-4}$ (note that using the EP test to determine parameters $\eta$ and $\beta$ assumes that there is no composition-induced EP violation).

Final results for all relativistic parameters obtained from the IfE (Institute f\"ur Erdmessung) analysis are shown in Table 2. 
The realistic errors are comparable with those obtained in other recent investigations, e.g.\ at JPL 
(see  Williams et al.\ 1996, 2004a, 2004b, 2005b).

\begin{table}[htb]
\caption{Determined values for the relativistic quantities and their realistic errors.}
\begin{center}
\begin{tabular}{|c|c|}
\hline
 Parameter    &  Results \\ \hline\hline
 diff.\ geod.\ prec.\ $\Omega_{\rm GP}$ - $\Omega_{\rm deSit}$ [''/cy] & $(6\pm 10)\cdot 10^{-3} $   \\ \hline
 metric par.\ $\gamma - 1$ & $(4 \pm 5) \cdot 10^{-3} $ \\ \hline
 metric par.\ $\beta - 1$ (direct)  & $(-2\pm 4) \cdot 10^{-3} $ \\ 
and from $\eta=4\beta-3-\gamma_{\rm Cassini}$ & $(1.5\pm 1.8) \cdot 10^{-4} $ \\ \hline
 equiv.\ principle par.\ $\eta$  & $(6 \pm 7) \cdot 10^{-4} $ \\ \hline
 time var.\ grav.\ const.\ $\dot G / G$ [yr$^{-1}$] & $(6\pm 8)\cdot 10^{-13} $  \\ \hline
 Yuk.\ coupl.\ const.\ $\alpha_{\lambda = 4\cdot 10^5\, {\rm km}}$
 & $(3\pm 2) \cdot 10^{-11} $ \\ \hline
 spec.\ relativi.\ $\zeta_1 - \zeta_0 - 1$  & $(-5\pm 12) \cdot 10^{-5} $
 \\ \hline
 infl.\ of dark matter $\delta g_{\rm galactic}$ [cm/s$^2$] & $(4\pm 4) \cdot 10^{-14} $
 \\ \hline
 `preferred frame' effect $\alpha_1$  & $(-7\pm 9) \cdot 10^{-5} $
 \\ \hline
 `preferred frame' effect $\alpha_2$  & $(1.8 \pm 2.5) \cdot 10^{-5} $
 \\ \hline

\end{tabular}
\end{center}
\end{table}

\section{Further Applications}

In addition to the relativistic phenomena mentioned above, more effects related to lunar physics, geosciences, and 
geodesy  can be investigated. The following items are of special interest:
\begin{enumerate}

\item Celestial reference frame: A dynamical realization of the International Celestial Reference System (ICRS)
 by the lunar orbit is obtained ($\sigma$~=~0.001") 
from LLR data. This can  be compared and analyzed with respect to the kinematical ICRS from VLBI. Here, the very good long-term stability of the orbit is of great advantage.  

\item 
Terrestrial reference frame: The results for the station coordinates and velocities, which are estimated simultaneously in the standard solution, contribute to the realization of the international terrestrial reference frame, e.g.\ to the
last one, the ITRF2000.

\item Earth rotation: LLR contributes, among others, to the determination of long-term nutation parameters, where again the stable, highly accurate orbit and the lack of non-conservative forces from atmosphere (which affect satellite orbits substantially) is very convenient. Additionally UT0 and VOL (variation of latitude) values are computed (e.g. Dickey et al. 1985), which stabilize the combined EOP series, especially in the 1970s when no good data from other space geodetic techniques were available. The precession rate is another example in this respect. The present accuracy of the long-term nutation coefficients and precession rate fits well with the VLBI solutions (within the present error bars), see (e.g. Williams et al. 2005a,c)

\item Relativity: In addition to the use of LLR in the more 'classical' geodetic areas, the dedicated investigation of Einstein's theory of relativity is of special interest.  With an improved accuracy the investigation of further effects 
(e.g.\ the Lense-Thirring precession) or those of alternative theories become possible.

\item Lunar physics: By the determination of the libration angles of the Moon, LLR gives access to underlying 
processes affecting lunar rotation (e.g.\ Moon's core, dissipation), cf. Williams et al.\ (2005a). A better 
distribution of the retro-reflectors on the Moon (see Fig.\ 7)  would be very helpful.

\item Selenocentric reference frame: The determination of a selenocentric reference frame, the combination with high-resolution images and the establishment of a better geodetic network on the Moon is a further big item, which then allows accurate lunar mapping. The LLR frame is used as reference in many application, e.g. to derive gravity models of the Moon (e.g. Konopliv et al. 2001).

 \item Earth-Moon dynamics: The mass of the Earth-Moon system, the lunar tidal acceleration, 
 possible geocenter variations and related processes as well as further effects can be investigated
 in detail. 
 
\item Time scales: The lunar orbit can also be considered as a long-term stable clock so that LLR can be 
used for the independent realization of time scales. Those features shall be addressed in the future.
\end{enumerate}

\begin{figure}[tbh]
  \begin{center}
  \epsfig{width=7.4cm,     file=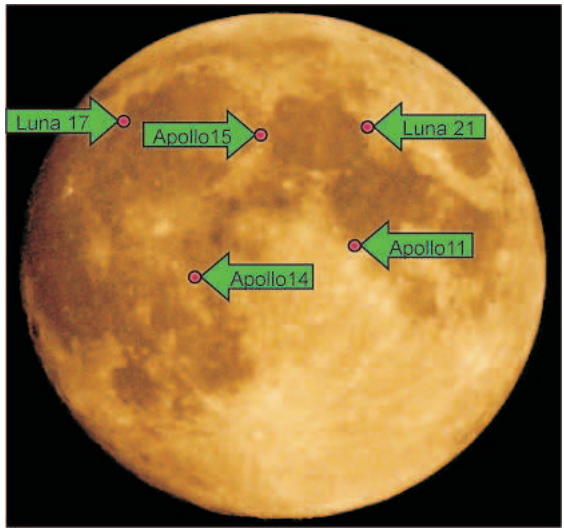}
  \end{center}
  \caption{Distribution of retro-reflectors on the Moon surface.} \label{refmoon}
 \end{figure}

To use the full potential of Lunar Laser Ranging, the theoretical models as well as the measurements require 
optimization. Using the 3.5 m telescope at the new Apollo site in New Mexico, USA, millimeter ranging becomes possible. 
To allow the determination of the various quantities of the LLR solution with a total gain of resolution of one order 
of magnitude, the models have to be up-dated according to the IERS conventions 2003, and  made compatible with the IAU 2000 resolutions. 
This requires, e.g., to better model
\begin{itemize}
\item higher degrees of the gravity fields of Earth and Moon and their couplings;
\item the effect of the asteroids (up to 1000);
\item relativistically consistent torques in the rotational equations of the Moon; 
\item relativistic spin-orbit couplings;
\item torques caused by other planets like Jupiter;
\item the lunar tidal acceleration with more periods (diurnal and semi-diurnal); 
\item ocean and atmospheric loading by updating the corresponding subroutines;
\item nutation using the recommended IAU model; 
\item the tidal deformation of Earth and Moon;
\item Moon's interior (e.g.\ solid inner core) and its coupling to the Earth-Moon dynamics.
\end{itemize}
Besides modeling, the overall LLR processing shall be optimized. The best strategy for the data fitting procedure 
needs to be explored for (highly) correlated parameters.

Finally LLR should be prepared for a renaissance of lunar missions where transponders (e.g. Degnan 2002) or new retro-reflectors may be deployed on the surface of the Moon which would enable many pure SLR stations to observe the Moon. NASA is planning to return to the Moon by 2008 with Lunar Reconnaissance Orbiter (LRO, 2005), and later with robotic landers, and then with astronauts in the middle of the next decade.  The primary focus of these planned missions will be lunar exploration and preparation for trips to Mars, but they will also provide opportunities for science, particularly if new reflectors are placed at more widely separated locations than the present configuration (see Fig.~7). New installations on the Moon would give stronger determinations of lunar rotation and tides. New reflectors on the Moon would provide additional accurate surface positions for cartographic control (Williams et al. 2005b), would also aid navigation of surface vehicles or spacecraft at the Moon, and they also would contribute significantly to research in fundamental and gravitational physics, LLR-derived ephemeris and lunar rotation. Moreover in the case of co-location of microwave transponders, the connection to the VLBI system may become possible which will open a wide range of further activities such as frame ties.

\section{Conclusions}
For the IERS, LLR has contributed to the realization of the 
International Terrestrial  Reference Frame ITRF2000 and to combined solutions of Earth Orientation 
Parameters.

Additionally, LLR has become a technique for measuring a variety of relativistic gravity parameters with unsurpassed precision. No definitive violation of the predictions from general relativity are found.  Both the weak and strong forms of the EP are verified, while strong empirical limitations on any inverse square law violation, time variation of $G$, and preferred frame effects are also obtained. 

LLR continues as an active program, and it can remain as one of the most important tools for testing Einstein's general relativity theory of gravitation if appropriate observations strategies are adopted and if the basic LLR model is further extended and improved down to the millimeter level of accuracy. 
 
Additional ranging devices on the Moon would have benefits for lunar science, fundamental physics, control networks for surface mapping, and navigation. Demonstration of active devices
would prepare the way for very accurate ranging to Mars and other solar system bodies.

\acknowledgments
{Current LLR data is collected, archived and distributed under the auspices of the International Laser Ranging Service (ILRS).  All former and current LLR data is electronically accessible through the European Data Center (EDC) in Munich, Germany and the Crustal Dynamics Data Information Service (CDDIS) in Greenbelt, Maryland.  The following web-site can be queried for further information: {\tt http://ilrs.gsfc.nasa.gov}.  We also acknowledge with thanks, that the more than 35 years of LLR data, used in these analyses, have been
obtained under the efforts of personnel at the Observatoire de la Cote
d¹Azur, in France, the LURE Observatory in Maui, Hawaii, and the McDonald
Observatory in Texas.

A portion of the research described in this paper was carried out at the Jet Propulsion Laboratory of the California Institute of Technology, under a contract with the National Aeronautics and Space Administration.}

\end{document}